\documentclass[reprint,amsmath,amssymb,aps,nofootinbib]{revtex4-2}
\usepackage{graphicx}
\graphicspath{ {images/} }
\usepackage{epstopdf}
\usepackage{hyperref}

\usepackage{color}
\definecolor{green}{rgb}{0,0.5,0}
\usepackage{xcolor}
\usepackage{textcomp}

\newcommand{\red}{\textcolor{black}} 

\newcommand{\blue}{\textcolor{black}} 
\usepackage[normalem]{ulem} % This package is for \sout{}. You must remove it before submitting the paper, to avoid issues with the submission process

\begin{document}

\title{\red{Non-equilibrium and first-passage properties of generalized Ornstein-Uhlenbeck processes under stochastic resetting}}

\author{Anna S. Bodrova$^{1}$}
\author{S. I. Serdyukov $^{2}$}

\affiliation{$^{1}$Moscow Institute of Electronics and Mathematics, National Research University Higher School of Economics, 123458, Moscow, Russia}

\affiliation{$^{2}$ Chemistry Department, M.V.Lomonosov Moscow State University, 119991 Moscow, Russia }

\date{\today}

\begin{abstract} 
\red{We study the non-equilibrium spatial properties of a generalized Ornstein-Uhlenbeck process in the presence of resetting dynamics in one dimension. To this end, we employ a recently developed extension of the Onsager-Machlup fluctuation theory \cite{Smain, Spub} to study the underlying process and then use renewal formalism to obtain the statistical properties of the resetting induced process. In particular, we compute the mean-squared displacement and probability density function in the non-equilibrium steady state and highlight the relaxation behavior of the properties towards the steady state. Next, we study the first passage properties of this process, confined to an interval. For instance, we compute MFPT and show that it can be further reduced by resetting. } \end{abstract}

%\keywords{Suggested keywords}%Use showkeys class option if keyword display desired
\maketitle

%\tableofcontents

\section{Introduction} 

%\textit{A process with resetting gets interrupted at certain points and starts anew. Resetting has been widely studied in the recent time and has numerous applications \cite{reviewres, reviewresgupta}. }

Processes with resetting can be observed in various fields \cite{reviewres, reviewresgupta}, such as computer algorithms \cite{computerscience,lor}, economics \cite{ecotax,eco1, eco2,ecoin}, psychology \cite{psi1,psi2}, and biology, including  enzyme-catalyzed reactions, described in terms of Michaelis-Menten kinetics \cite{chemistry,bio1,bio2,bio3}, transcription \cite{rna}, and animal mobility \cite{animals}. A similar concept was introduced in the previous century for catastrophic events \cite{cat1, cat2, cat3}. 

The term "resetting" was first used for particles performing  Brownian motion \cite{PRLinitial} and instantly returning to their starting point at random times. It has been shown that after a complicated time evolution, the probability density function (PDF) of a Brownian particle attains a non-equilibrium stationary state (NESS) \cite{PREmaj}. The fluctuation dissipation relation for Brownian motion with resetting was derived \cite{SokolovPRL2023}, and the ergodic properties of the resetting processes were investigated in \cite{eli, st}.

Different types of return processes have been considered, such as instantaneous jump to the starting point \cite{PRLinitial}, partial resetting \cite{partial}, and return at constant velocity \cite{shlomi, shlomi1,shlomi2,campos,Annasmooth,Annasmooth2d, radice} or  under the action of an external potential \cite{pot1, pot2, poti,exp1, exp2, exp3}. Resetting can significantly accelerate the search process \cite{SokChech,shlomimfpt,shlomi,palprasad}. It was shown that it is possible to determine an optimal restart rate that can minimize the mean first passage time.

Various processes with stochastic resetting have been studied previously, such as continuous time random walks \cite{Annactrw,MV2013,MC2016,Sh2017,ctrw,ctrwres}, L\'evy flights \cite{levy1,levy2}, L\'evy walks \cite{china}, heterogeneous diffusion processes \cite{hetero, fbm}, fractional Brownian motion \cite{fbm}, granular gases \cite{Anna25}, geometric Brownian motion \cite{gbm,ecoin, gbmresralf}, \blue{Ornstein-Uhlenbeck process \cite{OU2015, OU2023}}, scaled Brownian motion \cite{Annare,Annanonre}, and resetting on networks \cite{net}. 

In the present article, we discuss the influence of stochastic resetting on a new type of motion described in terms of the generalized Ornstein-Uhlenbeck distribution. %\cite{Ser25,Smain, Spub}. 
In Section II, we review the properties of this process published elsewhere \cite{Ser25,Smain, Spub, SerNS}. In Section III, we discuss the influence of resetting on the PDF, mean-squared displacement (MSD), and mean first passage times. \red{We derive the criteria for resetting either to favor or prohibit the search process.} Finally, in Section IV, we present our conclusions.

\section{Model} 
\subsection{Ornstein-Uhlenbeck process}
In this Section we consider the evolution of Brownian motion governed by the Langevin equation 
\begin{equation}
\frac{dv}{dt}=-\kappa v+\sqrt{2\varkappa}\eta(t), 
\label{even}
\end{equation} 
where $v=dx/dt$ is the velocity of the particle, $x$ is the coordinate of the particle, 
$\varkappa$ is the diffusion coefficient, $\kappa$ is the constraint coefficient, 
$t$ is the time, $\eta(t)$ is white noise. The white noise satisfies the following conditions: 
$$
\langle\eta(t)\rangle=0, \quad\langle\eta(t)\eta(t')\rangle=\delta(t-t'),
$$ 
where $\delta(t-t')$ is the delta-function. 

Furthermore, we consider the variational formulation of the fluctuation theory proposed by Onsager and Machlup \cite{Ons}.
%We proceed based on the variational fluctuation theory proposed by Onsager and Machlup \cite{Ons}.
%\blue{[Anna: the first sentence is a bit broken, not sure how to fix]}. 
Onsager and Machlup considered the deviation of a fluctuating system from equilibrium or a steady state described by a variable $\varphi(\tau)$, which satisfies the linear phenomenological equation 
 $-\dot\varphi=\kappa\varphi$, where $\dot\varphi=d\varphi/d\tau$, $0\leq\tau\leq t$ is the time.
%where $t$ is the time, 
%$\kappa$ is the phenomenological coefficient. 
Within the framework of the linear theory, the Lagrangian ${\cal L}(\varphi,\dot\varphi)$ 
and the corresponding functional $\psi[\varphi(\tau)]$ have the form 
\begin{eqnarray}
&&{\cal L}(\varphi,\dot\varphi)=\frac12(\dot\varphi+\kappa\varphi)^2 \label{lagr9}\\
&&\psi[\varphi]=\int_0^t{\cal L}(\varphi,\dot\varphi)\,d\tau =
\frac12\int_{0}^{t}(\dot\varphi+\kappa\varphi)^2 \,d\tau .
\label{lagr2}
\end{eqnarray}
From the condition 
\begin{equation}
\delta\psi=\int_0^t \delta{\cal L}\,d\tau=0
\end{equation} 
the Euler-Lagrange equation can be obtained:
\begin{equation}
\frac{d}{d\tau} \frac{\partial{\cal L}}{\partial \dot\varphi} -
\frac{\partial {\cal L}}{\partial \varphi} = 0 \,.
\label{eleven12}
\end{equation} 
It can be further reduced to the simpler equation 
\begin{equation}
\frac{d^2\varphi}{d\tau^2}- \kappa^2\varphi=0.
\label{e12}
\end{equation} 
%from where we obtain the solution (at conditions $\varphi(0)=x_0$, $\varphi(t)=x$):
%Solution of Eq.~(\ref{e12}) leads to the function (extremal):
with a solution (extremal)
\begin{equation}\varphi(\tau, t)=\frac{(ye^{\kappa t}-x)e^{-\kappa\tau}+(x-ye^{-\kappa t})e^{\kappa\tau} } { e^{\kappa t}- e^{-\kappa t}}, \label{tyu1}\end{equation} 
%\begin{equation}\varphi(\tau, t,x,y)=\frac{(ye^{\kappa t}-x)e^{-\kappa\tau}+(x-ye^{-\kappa t})e^{\kappa\tau} } { e^{\kappa t}- e^{-\kappa t}}, \label{tyu1}\end{equation} 
%where $\varphi(t)=x$, $\varphi(0)=y$ is the initial coordinate of the particle.
where $\varphi(0,t)=y$ is the initial coordinate of the particle and $\varphi(t,t)=x$.
From Eqs.~(\ref{lagr9}), (\ref{lagr2}), and (\ref{tyu1}), we obtain Lagrangian 
${\cal L}(\tau,t,x,y)$  
and its functional minimum $\psi(t,x)$: 
\begin{eqnarray}
&&{\cal L}(\tau, t,x,y)=
\frac{ 2\kappa^2 e^{2\kappa\tau}(x-ye^{-\kappa t})^2}{e^{2\kappa t}(1-e^{-2\kappa t})^2}, 
\\
&&\psi(t,x,y)=\int_0^t{\cal L}(\tau, t,x,y)\,d\tau = \frac{\kappa(x-ye^{-\kappa t})^2}{1-e^{-2\kappa t}}.
\label{lagg2}
\end{eqnarray} 

\begin{figure}\centerline{\includegraphics[width=0.5\textwidth]{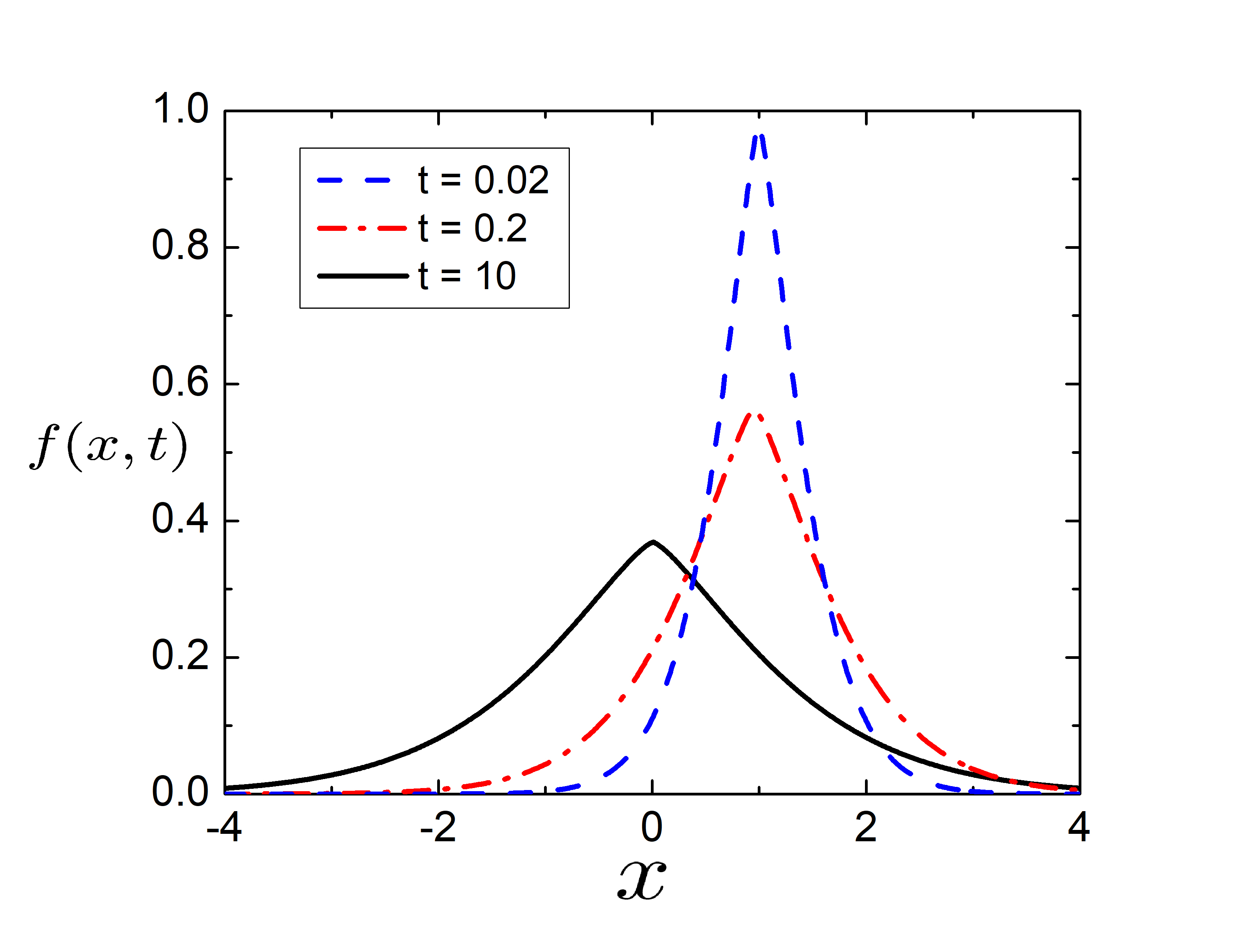}}\caption{The evolution of PDF (Eq.~\ref{mainpdf}) for $\mu=\frac{4}{3}$, $\nu=4$, $\varkappa=\frac{1}{2}$, $\kappa=\frac{1}{4}$, $x_0=1$. At $t = 10$ the stationary state given by Eq.~(\ref{pdfstat}) is obtained.} \label{Gpdf}
\end{figure}

Onsager and Machlup introduced functional representation of the corresponding probability density function~\cite{Ons}
\begin{equation}
\label{40f2}
f= f_0\exp\left(-\frac{\psi}{2\varkappa} \right) .
\end{equation}
Substitution of Eq.~(\ref{lagg2}) into Eq.~(\ref{40f2}) yields the probability density function of the Ornstein-Uhlenbeck random process:  
\begin{equation}
\label{40fee2}
f(t,x|y)= \sqrt{\frac{\kappa}{2\pi\varkappa (1-e^{-2\kappa t})}}
\exp\left\{ -\frac{\kappa (x-y e^{-\kappa t})^2 }
              {2{\varkappa}(1-e^{-2\kappa t})}  \right\} . 
\end{equation}

This function describes the evolution of a non-equilibrium system to an equilibrium 
or steady state ($t\to\infty$).
In the steady-state, the PDF given by Eq.~(\ref{40fee2}), is reduced to 
\begin{equation}
\label{40fqq1}
f(x)= \sqrt{\frac{\kappa}{2\pi\varkappa}}\exp\left(-\frac{{\kappa}x^2}{2{\varkappa}} \right). 
\end{equation}
%Here, $\varkappa$ is the diffusion coefficient. 
At $\kappa\to 0$ we obtain from Eq.~(\ref{40fee2}):
\begin{equation}
\label{4ee2}
\red{f(t,x|y)}= \frac{1}{2\sqrt{\pi\varkappa t}}\exp\left\{ -\frac{(x-y)^2 } {4\varkappa t} \right\} . 
\end{equation}

The probability density function of the Ornstein-Uhlenbeck random process, given by Eq.~(\ref{40fee2}), can be obtained as a solution to the well-known Fokker-Planck equation 
\begin{equation}
\label{fk22}
\frac{\partial f}{\partial t}=
  \kappa \frac{\partial}{\partial x}(x f)+\varkappa\frac{\partial^2 f}{\partial x^2}\, ,
\end{equation}
%where $\kappa$ is the potential strength. 
which reduces to the diffusion equation  at $\kappa=0$:
\begin{equation}
\frac{\partial f}{\partial t}=\varkappa\frac{\partial^2 f}{\partial x^2} .
\end{equation}

\red{\subsection{Generalized Ornstein-Uhlenbeck process}}

The theory of Onsager and Machlup was generalized in \cite{Smain}, where the Lagrangian 
and functional minima were represented as
%\label{qua2}
\begin{eqnarray}\label{lagrr}
&& {\cal L}(\varphi, \dot\varphi) = \frac1{\mu} |\dot\varphi+\kappa\varphi|^{\mu} ,\label{bnk62}\\ 
&&\psi[\varphi]=\int_0^t {\cal L}(\varphi,\dot\varphi)\,d\tau=
\frac1{\mu} \int^{t}_{0} |\dot\varphi+\kappa\varphi|^{\mu}d\tau. 
\label{bnt35}
\end{eqnarray} 
Introducing the Lagrangian (Eq.~\ref{lagrr}) into the Euler-Lagrange equation (\ref{eleven12}), we obtain the following second-order equation:
\begin{equation}
(\mu-1)\frac{d^2\varphi}{d\tau^2}+(\mu-2)\kappa\frac{d\varphi}{d\tau} - \kappa^2\varphi=0.
\label{el2}
\end{equation} 
The solution to Eq.~(\ref{el2}) is
\begin{equation}\varphi(\tau, t)=\frac{(ye^{\frac{\kappa t}{\mu-1}}-x)e^{-\kappa\tau}+(x-ye^{-\kappa t})e^{\frac{\kappa\tau}{\mu-1}} } { e^{\frac{\kappa t}{\mu-1}}- e^{-\kappa t}}. \label{tyu}\end{equation} 
In the classical case ($\mu=2$), Eq.~(\ref{el2}) can be reduced to (Eq.~\ref{e12}).
%\begin{equation}\varphi(\tau, t,x,y)=\frac{(ye^{\frac{\kappa t}{\mu-1}}-x)e^{-\kappa\tau}+(x-ye^{-\kappa t})e^{\frac{\kappa\tau}{\mu-1}} } { e^{\frac{\kappa t}{\mu-1}}- e^{-\kappa t}}. \label{tyu}\end{equation} 

Generalized theory permits us to derive the higher-order Lagrangian ${\cal L}(\tau,t,x,y)$ 
and functional minimum $\psi(t,x,y)$:
\begin{eqnarray}\label{Lbnt62}
&&{\cal L}(\tau,t,x,y) = \frac{(\nu\kappa)^{\mu} e^{\nu\kappa \tau} |x-y e^{-\kappa t}|^{\mu}} 
{ \mu e^{\nu\kappa t} (1-e^{-\nu \kappa t })^{\mu}}, \\ 
&&\psi(t,x,y) = 
\frac{(\nu\kappa)^{\mu-1}|x-y e^{-\kappa t}|^{\mu}} {\mu(1-e^{-\nu \kappa t })^{\mu-1}}\label{bnt62}.
\end{eqnarray} 
Here, the shape parameters $\nu$ and $\mu$ are related by 
\begin{equation}\nonumber
\frac{1}{\mu}+\frac{1}{\nu}=1\,,\;\:\mu>1\,,\;\:\nu>1\,.
\end{equation}

\begin{figure}\centerline{\includegraphics[width=0.5\textwidth]{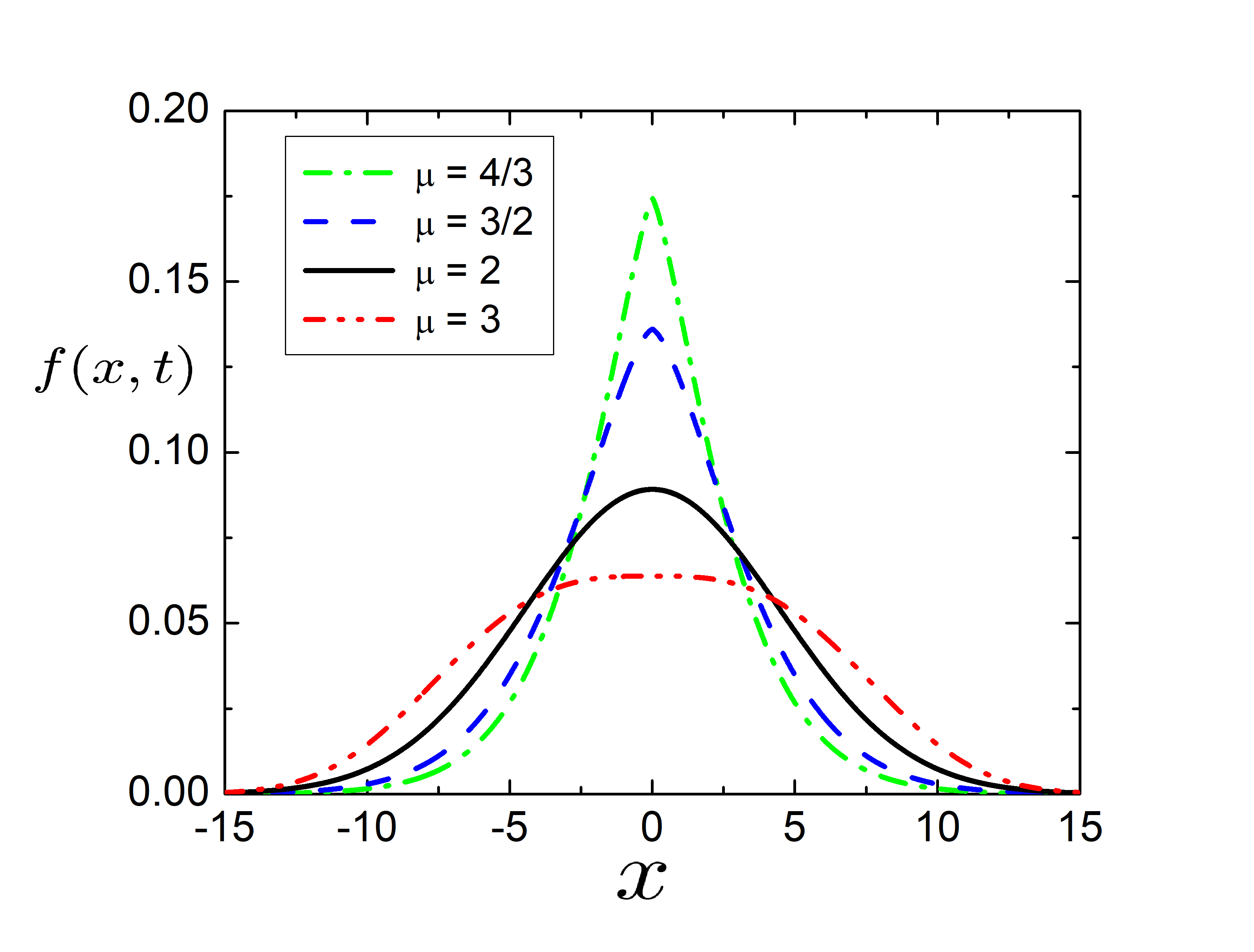}}\caption{\blue{PDF (Eq.~\ref{pdfsimple}) for $t=20$, $\mu=\frac{4}{3}, \frac{3}{2}, 2, 3$,  $\varkappa=\frac{1}{2}$.}} \label{Gpdffree}\end{figure}

The generalized PDF depends now on $\mu$ and $\nu$: 
%\footnote{In \cite{Smain}, $f(t,x)\sim\exp\left\{-\frac{\psi}{2\varkappa}\right\}$ has been used .}
\begin{equation}
f=f_0\exp\left\{-\frac{\psi}{(\nu\varkappa)^{\mu-1}}\right\} .
\label{b6}
\end{equation} 
Introducing Eq.~(\ref{bnt62}) into Eq.~(\ref{b6}), we obtain the generalized Ornstein-Uhlenbeck PDF %generalized Ornstein-Uhlenbeck type PDF
\begin{eqnarray}
\nonumber
f(t,x|y)=\frac{\left(\mu\kappa\right)^{\frac{1}{\nu}}}{2\Gamma\left(\frac{1}{\mu}\right)\varkappa^{\frac{1}{\nu}}\left(1-e^{-\nu\kappa t}\right)^{\frac{1}{\nu}}}\\
\times\exp\left(-\frac{\kappa^{\mu-1}|x-ye^{-\kappa t}|^{\mu}}{\mu\varkappa^{\mu-1}\left(1-e^{-\nu\kappa t}\right)^{\mu-1}}\right)\,.
\label{mainpdf}%\label{b6912}
\end{eqnarray}
This expression corresponds to the generalized Ornstein-Uhlenbeck distribution. Its evolution is shown in Fig. \ref{Gpdf}. Over longer time periods, the distribution becomes wider, and the peak value shifts to zero. The generalized Ornstein-Uhlenbeck distribution obeys a symmetrical shape and corresponds to a well-studied generalized normal distribution \cite{nadar,pog}. At $\mu=\nu=2$ we obtain the case considered in the theory of Onsager and Machlup \cite{Ons}.

The Langevin type equation in the general case is derived in \cite{Ser25}. 
In this case, instead of (Eq.~\ref{even}) we get
\begin{equation}
\frac{dv}{dt}=-\kappa v+(\nu!\,\varkappa)\zeta_{\nu}(t), 
\label{eve1}
\end{equation} 
with the noise $\zeta_{\nu}(t)$. 
Generally, a random process corresponding to (Eq.~\ref{eve1}), is non-Marcovian with a non-zero correlation time.

Let us set $y=0$ and consider the discrete form parameters $\mu=1+1/n$, $\nu=n+1$, $n=1,2,\dots$. The generalized Ornstein-Uhlenbeck type distribution (Eq.~\ref{mainpdf}) %takes the form
%\begin{eqnarray}\nonumberf(x,t)=\frac{  \left( 1+\frac1{n}\right) ^{\frac1{n+1}}  } {2\Gamma(\frac{n}{n+1}) } \left(\frac{\kappa}{\varkappa (1-e^{-(n+1)\kappa t}) } \right)^{\frac1{n+1}}\\\times   \exp \left\{ -\left(\frac{\kappa}{\varkappa (1-e^{-(n+1)\kappa t}) } \right)^{\frac1{n}}   \frac{|x|^{1+\frac1n}} {\mu}\right\}\,.\label{a9p2}\end{eqnarray}
can be considered as a solution to the differential equation 
\begin{eqnarray}\nonumber
\frac{\partial f(t,x)}{\partial t}=\kappa\frac{\partial}{\partial x}(xf(t,x))+(-1)^{n+1}\varkappa\\
\times\left( 
\frac{\partial}{\partial x} \frac{a_1}{x^{n}}+\dots+\frac{\partial^{n}}{\partial x^{n}}\frac{a_{n}}{x}+ 
\frac{\partial^{n+1}} {\partial x^{n+1}} \right)  f(t,x) 
\label{difeqw}
\end{eqnarray}
with solution
\begin{eqnarray}
\nonumber
f(t,x)=\frac{\left(\mu\kappa\right)^{\frac{1}{\nu}}}{2\Gamma\left(\frac{1}{\mu}\right)\varkappa^{\frac{1}{\nu}}\left(1-e^{-\nu\kappa t}\right)^{\frac{1}{\nu}}}\\
\times\exp\left(-\frac{\kappa^{\mu-1}|x|^{\mu}}{\mu\varkappa^{\mu-1}\left(1-e^{-\nu\kappa t}\right)^{\mu-1}}\right)\,.
\label{inpdf}
\end{eqnarray}

\red{For example, at} $n=3$, $\mu=\frac{4}{3}$, $\nu=4$ Eq.~(\ref{difeqw}) is reduced to a fourth-order equation \cite{Ser25}:
\begin{eqnarray}
\label{difeq4}
\frac{\partial f(t,x)}{\partial t}&=&\kappa \frac{\partial}{\partial x}\left(x f(t,x)\right)\\
&+&\varkappa\left(\frac{\partial}{\partial x}\frac{a_1}{x^3}+\frac{\partial^{2}}{\partial x^{2}}\frac{a_{2}}{x^2}+\frac{\partial^{3}}{\partial x^{3}}\frac{a_{3}}{x}+\frac{\partial^{4}}{\partial x^{4}}\right)f(x,t)\nonumber
\end{eqnarray}
with
\begin{equation}
a_1=-\frac{8}{9}\,;\;\; a_2=-\frac{13}{9};\;\;a_3=-1\,.
\end{equation}
In this case Eq.~(\ref{mainpdf}) takes the following form:
\begin{equation}
f(t,x)=\frac{\left(\frac43\kappa\right)^{\frac14}}{2\Gamma\left(\frac{3}{4}\right)
\varkappa^{\frac14}(1-e^{-4\varkappa t})^{\frac14}}
\exp\left(-\frac{\kappa^{\frac13}|x|^{\frac43}}{\frac43\varkappa^{\frac13}(1-e^{-4\varkappa t})^{\frac43}}
\right)\,.
\label{ainpdf}
\end{equation}
Thus, from the Fokker-Plank equation (Eq.~\ref{fk22}), we proceed to higher-order equations (Eqs.~\ref{difeqw}).

At $t\to\infty$ the PDF (Eq.~\ref{mainpdf}) tends to the stationary state:
\begin{equation}
f(t,x)=\frac{\left(\mu\kappa\right)^{\frac{1}{\nu}}}{2\Gamma\left(\frac{1}{\mu}\right)\varkappa^{\frac{1}{\nu}}}\exp\left(-\frac{\kappa^{\mu-1}|x|^{\mu}}{\mu\varkappa^{\mu-1}}\right)\,.
\label{pdfstat}
\end{equation}
At $\kappa\to 0$, corresponding to free motion without confinement, the PDF (Eq.~\ref{mainpdf}) becomes
\begin{equation}
f(t,x)=\frac{\mu^{\frac{1}{\nu}}}{2\Gamma\left(\frac{1}{\mu}\right)\left(\nu\varkappa t\right)^{\frac{1}{\nu}}}\exp\left(-\frac{|x|^{\mu}}{\mu\left(\nu\varkappa t\right)^{\mu-1} }\right)\,. \label{pdfsimple}
\end{equation}
The PDF, given by Eq.~(\ref{pdfsimple}), is presented in Fig.~\ref{Gpdffree} for different values of $\mu$, where $\mu=2$ corresponds to standard Brownian diffusion, and with decreasing of the parameter $\mu$ the distribution becomes sharper. For $\mu = 3$ the distribution is flatter than that for standard Brownian motion.\\

\subsection{Mean-squared displacement}

The MSD can be derived according to 
\begin{equation}\left< x^2(t) \right> = \int_{-\infty}^{\infty}x^2 f(t,x) dx\,,
\label{x2}
\end{equation}
Setting $y=0$ and performing the integration of Eq.~(\ref{mainpdf}), we get \cite{Smain}
\begin{equation}\left< x^2(t) \right> = \frac{\Gamma\left(\frac{3}{\mu}\right)\mu^{\frac{2}{\mu}}\varkappa^{\frac{2}{\nu}}}{\Gamma\left(\frac{1}{\mu}\right)}\left(\frac{1-e^{-\nu\kappa t}}{\kappa}\right)^{\frac{2}{\nu}}\,.
\label{x2full}
\end{equation}
At $t\to\infty$ the MSD tends to the constant value
\begin{equation}\left< x^2(t) \right> = \frac{\Gamma\left(\frac{3}{\mu}\right)\mu^{\frac{2}{\mu}}\varkappa^{\frac{2}{\nu}}}{\Gamma\left(\frac{1}{\mu}\right)\kappa^{\frac{2}{\nu}}}\,.\label{x2s}
\end{equation}
In the absence of an external potential $\kappa\to 0$ the MSD has a power-law dependence, typical for an anomalous diffusion process \cite{Ralfmain, Ralfdop}:
\begin{equation}\left< x^2(t) \right> = K_{\alpha}t^{\alpha},
\label{x2k}
\end{equation}
where the power law constant of the MSD depends on the shape parameters of the generalized normal distribution:
\begin{eqnarray}
\alpha&=&\frac{2}{\nu}=2-\frac{2}{\mu}\,.\label{anu}
\end{eqnarray} 
Here the prefactor is equal to
\begin{eqnarray}\label{Ka}
K_{\alpha}&=&\frac{\Gamma\left(\frac{3}{\mu}\right)}{\Gamma\left(\frac{1}{\mu}\right)}\mu^{\frac{2}{\mu}}\left(\nu\varkappa\right)^{\frac{2}{\nu}}
\end{eqnarray} 
%In this way, we obtain anomalous diffusion processes with  non-linear dependence of the MSD on time \cite{Ralfmain, Ralfdop}. 
 \blue{Superdiffusion is established when $\alpha>1$ or $\mu>2$. The parameter range $0<\alpha<1$ or $1<\mu<2$ corresponds to subdiffusion.} At $\mu=2$ the normal diffusion is obtained as follows
\begin{equation}\left< x^2(t) \right> = 2\varkappa t\,.
\end{equation}

\subsection{Higher order moments}

The fourth moment can be expressed as
\begin{equation}\left< x^4(t) \right> = \int_{-\infty}^{\infty}x^4 f(x,t) dx\,.
\end{equation}
The integration yields
\begin{equation}\left< x^4(t) \right> = \frac{\Gamma\left(\frac{5}{\mu}\right)\mu^{\frac{4}{\mu}}\varkappa^{\frac{4}{\nu}}}{\Gamma\left(\frac{1}{\mu}\right)}\left(\frac{1-e^{-\nu\kappa t}}{\kappa}\right)^{\frac{4}{\nu}}\,.
\label{x4}
\end{equation}
At $\kappa\to 0$ it becomes
\begin{equation}\left< x^4(t) \right> = \frac{\Gamma\left(\frac{5}{\mu}\right)\mu^{\frac{4}{\mu}}\left(\nu\varkappa\right)^{\frac{4}{\nu}}}{\Gamma\left(\frac{1}{\mu}\right)}t^{\frac{4}{\nu}}\,.
\label{x4s}
\end{equation}
The kurtosis is equal to
\begin{equation}
K=\frac{\left< x^4(t) \right>}{\left(\left< x^2(t) \right>\right)^2}=\frac{\Gamma\left(\frac{5}{\mu}\right)\Gamma\left(\frac{1}{\mu}\right)}{\left(\Gamma\left(\frac{3}{\mu}\right)\right)^2}\,.
\end{equation}
For $\mu=\frac{4}{3}$ the kurtosis $K=4.22$, which corresponds to a sharper distribution than the normal distribution. 

The odd moments of the generalized Ornstein-Uhlenbeck distribution (Eq.~\ref{mainpdf}) with $y=0$ are equal to zero:
\begin{equation}\left< x^{2k+1}(t) \right> = 0\,,\;\; \text{for}\quad k=0,1,... 
\end{equation}
The generalized equation for even moments reads:
\begin{eqnarray}\left< x^{2k}(t) \right>  =  \int_{-\infty}^{\infty}x^{2k} f(x,t) dx \\
=\frac{\Gamma\left(\frac{2k+1}{\mu}\right)\mu^{\frac{2k}{\mu}}\varkappa^{\frac{2k}{\nu}}}{\Gamma\left(\frac{1}{\mu}\right)}\left(\frac{1-e^{-\nu\kappa t}}{\kappa}\right)^{\frac{2k}{\nu}}\\
\text{for}\quad  k=0,1,...\nonumber
\end{eqnarray}

\section{Resetting} 

Let us consider a particle returning to its initial position, located in the origin, at random times. We denote the PDF of the waiting times between two consecutive resetting events as $\psi(t)$. 
We consider exponential resetting, which corresponds to the Poisson process:
\begin{equation}
\psi (t) = r{e^{ - rt}}
\label{pdfexp}
\end{equation}
where $r$ is the constant rate of the resetting events.
The survival probability $\Psi(t)$ is defined as the probability that no resetting event occurs between zero and $t$,
\begin{equation}
\Psi (t) = 1 - \int\limits_0^t {\psi (t')dt'}  =  {e^{ - rt}}\,.
\label{surv}
\end{equation}

\subsection{Probability density function under resetting}

\begin{figure}\centerline{\includegraphics[width=0.5\textwidth]{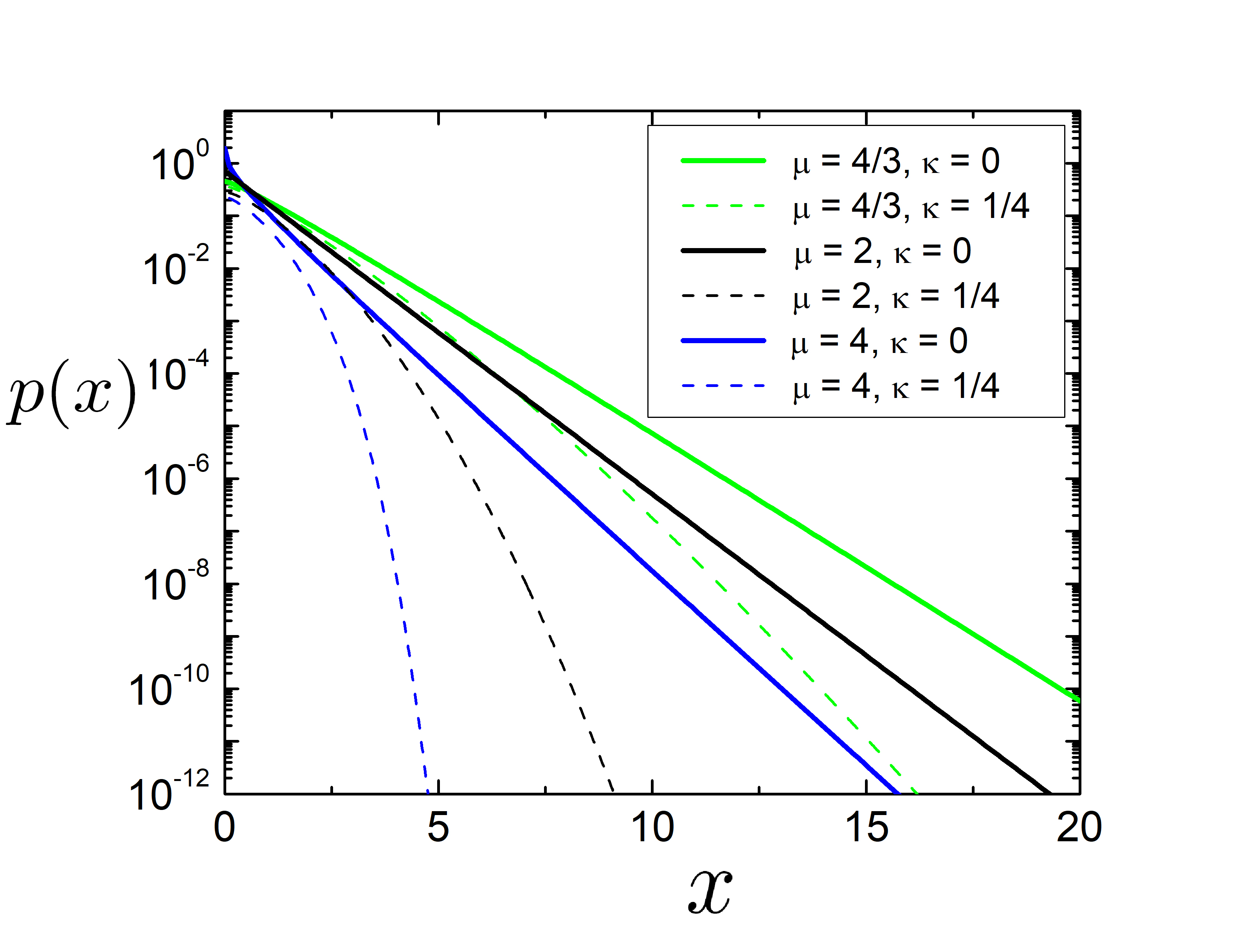}}\caption{\red{The NESS of the PDF with resetting for $r=1$, $\mu=\frac{4}{3},\,2,\,4$,  $\varkappa=\frac{1}{2}$ with confinement ($\kappa=\frac{1}{4}$, dashed lines) and without confinement ($\kappa=0$, solid lines) at $t=50$. With increasing of $\mu$ and in confinement the distribution becomes more narrow. PDF in confinement is obtained by numerical integration of Eq.~(\ref{eqprob2}). NESS for $\kappa=0$ is derived according to Eq.~(\ref{pdfres0}).}} \label{Gpdfres}
\end{figure}

The probability density for finding a particle at location $x$ at time $t$ with rate $r$ is given by 
\begin{equation}
p(x,t) = \Psi (t)f(x,t) + \int\limits_0^t dt^{\prime} r \Psi (t - t^{\prime})f(x,t-t^{\prime})\,.
\label{eqprob}
\end{equation}
Here, the first term accounts for the realizations where no resetting occurrs up to observation time $t$. The second term accounts for the case when the resetting event occurs at time $t^{\prime}$, and there are no resetting events between $t^{\prime}$ and $t$. 
By introducing the survival probability, Eq.~(\ref{surv})  into Eq.~(\ref{eqprob}), and using a new variable $\tau=t - t^{\prime}$, we obtain
\begin{equation}
p(x,t) = e^{-rt} f(x,t) + r\int\limits_0^t d\tau e^{-r\tau} f(x,\tau)\,.\label{eqproban}
\end{equation}
The first term may be safely neglected after a long time period
\begin{equation}
p(x,t) =r\int\limits_0^t d\tau e^{-r\tau} f(x,\tau)\,.
\label{eqprob1}
\end{equation}
For simplicity, we assume that the particle returns with resetting to the initial position, $y=0$. The PDF can be obtained by inserting Eq.~(\ref{mainpdf}) into Eq.~(\ref{eqprob1}).
\begin{eqnarray}\nonumber
&&p(x,t) =r\int\limits_0^t d\tau e^{-r\tau} \frac{\left(\mu\kappa\right)^{\frac{1}{\nu}}}{2\Gamma\left(\frac{1}{\mu}\right)\varkappa^{\frac{1}{\nu}}\left(1-e^{-\nu\kappa \tau}\right)^{\frac{1}{\nu}}}\\
&&\times\exp\left(-\frac{\kappa^{\mu-1}|x|^{\mu}}{\mu\varkappa^{\mu-1}\left(1-e^{-\nu\kappa \tau}\right)^{\mu-1}}\right)\,.
\label{eqprob2}
\end{eqnarray}
The integral was calculated numerically using Wolfram Mathematica 12.0. The results are shown in Fig.~\ref{Gpdfres}.

When $\kappa=0$, which corresponds to motion without confinement, the PDF can be derived analytically. Using Eq.~(\ref{pdfsimple}) and Eq.~(\ref{eqprob1}), we get 
\begin{eqnarray}
p(x,t) =\frac{r\mu^{\frac{1}{\nu}}}{2\Gamma\left(\frac{1}{\mu}\right)\left(\nu\varkappa\right)^{\frac{1}{\nu}}}\int_0^{t}\tau^{-\frac{1}{\nu}}e^{-\xi\left(\tau\right)}d\tau\,.\label{pdfresx}
\end{eqnarray}
Here we introduce the function
\begin{equation}
\xi(\tau)=r\tau+\frac{\tau^{1-\mu}|x|^{\mu}}{\mu\left(\nu\varkappa\right)^{\mu-1}}\,.
\end{equation}
It attains the minimum at $\tau=\tau_0$, which can be found from the condition 
\begin{equation}
\xi^{\prime}(\tau_0)=r+\left(1-\mu\right)\frac{\tau_0^{-\mu}|x|^{\mu}}{\mu\left(\nu\varkappa\right)^{\mu-1}}=0
\end{equation}
and equal to
\begin{eqnarray}
\tau_0=\frac{|x|}{\nu\varkappa^{\frac{1}{\nu}}r^{\frac{1}{\mu}}}\,.
\end{eqnarray}
The extreme value of this function at $\tau=\tau_0$ is
\begin{eqnarray}\label{v0}
\xi(\tau_0)=\left(\frac{r}{\varkappa}\right)^{\frac{1}{\nu}}|x|\,.
\end{eqnarray}
The second derivative of $\xi(\tau)$ at $\tau = \tau_0$ can be calculated as
\begin{eqnarray}\label{v00}
\xi^{\prime\prime}(\tau_0)=\left(\mu-1\right)\frac{\tau_0^{-\mu-1}|x|^{\mu}}{\left(\nu\varkappa\right)^{\mu-1}}=\frac{\mu\nu r^{1+\frac{1}{\mu}}\varkappa^{\frac{1}{\nu}}}{|x|}\ge 0\,.
\end{eqnarray}
Next, we apply a saddle-point approximation. We perform a Taylor expansion for $\xi(\tau)$
\begin{equation}\label{tay}
\xi(\tau)=\xi\left(\tau_0\right)+\frac{1}{2}\left(\tau-\tau_0\right)^2\xi^{\prime\prime}(\tau_0)+...
\end{equation}
and introduce it into the PDF given by Eq.~(\ref{pdfresx}). Let us assume that $1\ll\tau_0\ll t$ or 
\begin{eqnarray}\nonumber|x|\gg\nu\varkappa^{\frac{1}{\nu}}r^{\frac{1}{\mu}}\;;\;\;\;\;t\gg\frac{|x|}{\nu\varkappa^{\frac{1}{\nu}}r^{\frac{1}{\mu}}}\,.\end{eqnarray}
In this case, the upper and lower integration limits tend to $\pm \infty$ and the PDF can be written as
\begin{eqnarray}\nonumber
&&p(x,t)=\frac{r\mu^{\frac{1}{\nu}}}{2\Gamma\left(\frac{1}{\mu}\right)\left(\nu\varkappa\tau_0\right)^{\frac{1}{\nu}}}\int_{-\infty}^{\infty}e^{-\xi(\tau_0)-\frac{\left(\tau-\tau_0\right)^2}{2}\xi^{\prime\prime}(\tau_0)}d\tau\\
&&=\frac{\sqrt{2\pi}r\mu^{\frac{1}{\nu}}}{2\Gamma\left(\frac{1}{\mu}\right)\sqrt{\xi^{\prime\prime}(\tau_0)}\left(\nu\varkappa\tau_0\right)^{\frac{1}{\nu}}}\exp\left(-\xi\left(\tau_0\right)\right)\,.
\label{pdfres0x}
\end{eqnarray}
Introducing Eqs.~(\ref{v0}-\ref{v00}) into Eq.~(\ref{pdfres0x}), we obtain the steady-state value of the PDF:
\begin{eqnarray}
p(x)=\frac{\sqrt{2\pi}r^{\frac{1}{2\nu}}|x|^{\frac{1}{2}-\frac{1}{\nu}}\varkappa^{\frac{1}{\nu^2}-\frac{3}{2\nu}}}{2\Gamma\left(\frac{1}{\mu}\right)\nu^{\frac{1}{2}}\mu^{\frac{1}{2}-\frac{1}{\nu}}}\exp\left(-\left(\frac{r}{\varkappa}\right)^{\frac{1}{\nu}}|x|\right).\qquad
\label{pdfres0}
\end{eqnarray}
The NESS for different values of parameter $\mu$ and $\kappa$ is presented in Fig.~\ref{Gpdfres}. It becomes narrower with an increase in the shape parameter $\mu$ and the potential strength $\kappa$. 

At $t<\tau_0$ the minimum can not be attained during the observation period. The most significant impact on the integrand occurs at $\tau\approx t$. In this case, the PDF attains the form
\begin{equation}
p(x,t)\sim\left(1+r\right)\frac{\mu^{\frac{1}{\nu}}}{2\Gamma\left(\frac{1}{\mu}\right)\left(\nu\varkappa t\right)^{\frac{1}{\nu}}}\exp\left(-rt-\frac{|x|^{\mu}}{\mu\left(\nu\varkappa t\right)^{\mu-1} }\right)\,.\label{pdfsmallres}
\end{equation}
%\blue{It corresponds to trajectories, which have undergone no resettings up to time $t$ \cite{PREmaj}.}

\begin{figure}\centerline{\includegraphics[width=0.5\textwidth]{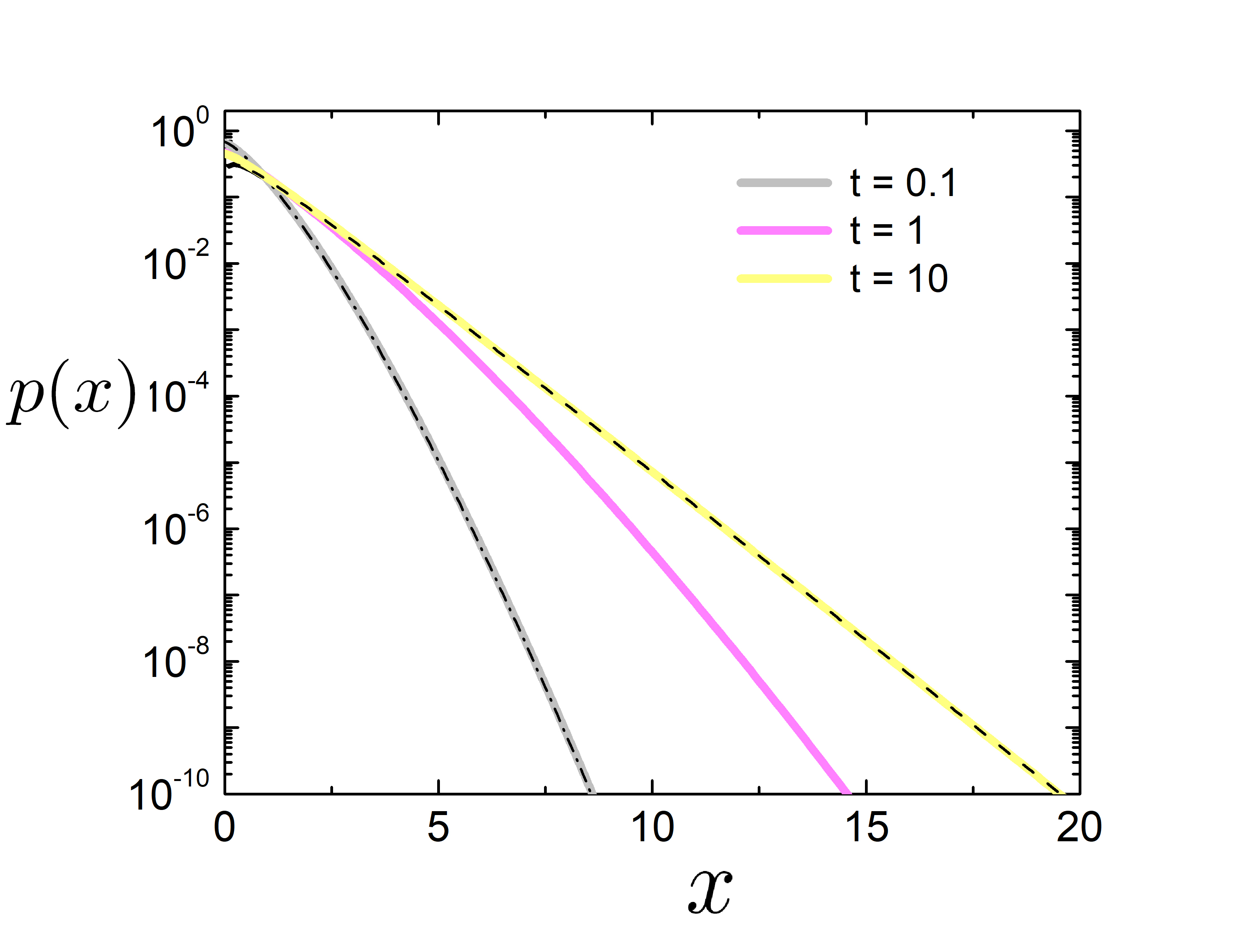}}\caption{\red{The PDF with resetting for $r=1$, $\mu=\frac{4}{3}$, $\nu=4$, $\varkappa=\frac{1}{2}$, $\kappa=0$ at $t=0.1,\,1,\,3,\,10$. Black dashed line shows PDF in the steady state (Eq.~\ref{pdfres0}). Black dashed-dotted line depicts PDF free motion without resetting (Eq.~\ref{pdfsimple}). Solid yellow, green, magenta and gray lines correspond to numerical integration of Eq.~(\ref{eqproban}) with propagator given by Eq.~(\ref{pdfsimple}) and show the relaxation towards the steady state.}} \label{Gpdfmaj}
\end{figure}

The PDF at different times is shown in Fig.~\ref{Gpdfmaj}. At short times, the number of resetting events is negligible and the PDF is practically the same as that without resetting (Eq.~\ref{pdfsimple}). At long times, the PDF tends toward the NESS, given by Eq.~(\ref{pdfres0}). At intermediate times, the NESS is established in a core region around the resetting center, whereas outside the core region, the system is transient. \red{The stationary front grows over time at a constant velocity $\nu\varkappa^{\frac{1}{\nu}}r^{\frac{1}{\mu}}$.} This phenomenon was first observed by Majumdar et al. for Brownian motion, fractional Brownian motion, and fluctuating interfaces \cite{PREmaj}. It was then investigated by Pal et al. for Brownian motion with time-dependent resetting \cite{Paltime}, and by Singh et al. for diffusion in confining potentials \cite{Singhtime}. 

%This analytical estimation is shown at Fig.~\ref{Gpdfres} as black dashed line. It can be compared with numerical derivation of PDF, obtained by inserting (Eq.~\ref{pdfsimple}) into (Eq.~\ref{eqproban}) and performing numerical integration using Mathematica. The result corresponds to red solid line at Fig.~\ref{Gpdfres}. It may be seen that the analytical estimation practically coincide with the result of numerical integration at $t = 50$.

 %At $t\to\infty$ it tends to the stationary distribution function (\ref{pdfstat}).

\subsection{Mean-squared displacement under resetting}

Multiplying Eq. (\ref{eqprob}) by $x^2$ and performing integration over $x$, we obtain the following equation for the MSD of the particles:
\begin{eqnarray}\nonumber
\left\langle x^2(t)\right\rangle = \left\langle x^2(t)\right\rangle \Psi(t) + \int_0^t dt^{\prime}r\Psi(t-t^{\prime})\left\langle x^2(t-t^{\prime})\right\rangle\,.\\
\label{x2beg}
\end{eqnarray}

At long times $t\to\infty$ the first term may be neglected and the MSD becomes
\begin{eqnarray}\label{x2long}
\left\langle x^2(t)\right\rangle =\int_0^t dt^{\prime} r\Psi(t-t^{\prime})\left\langle x^2(t-t^{\prime})\right\rangle\,.
\end{eqnarray}

\begin{figure}\centerline{\includegraphics[width=0.5\textwidth]{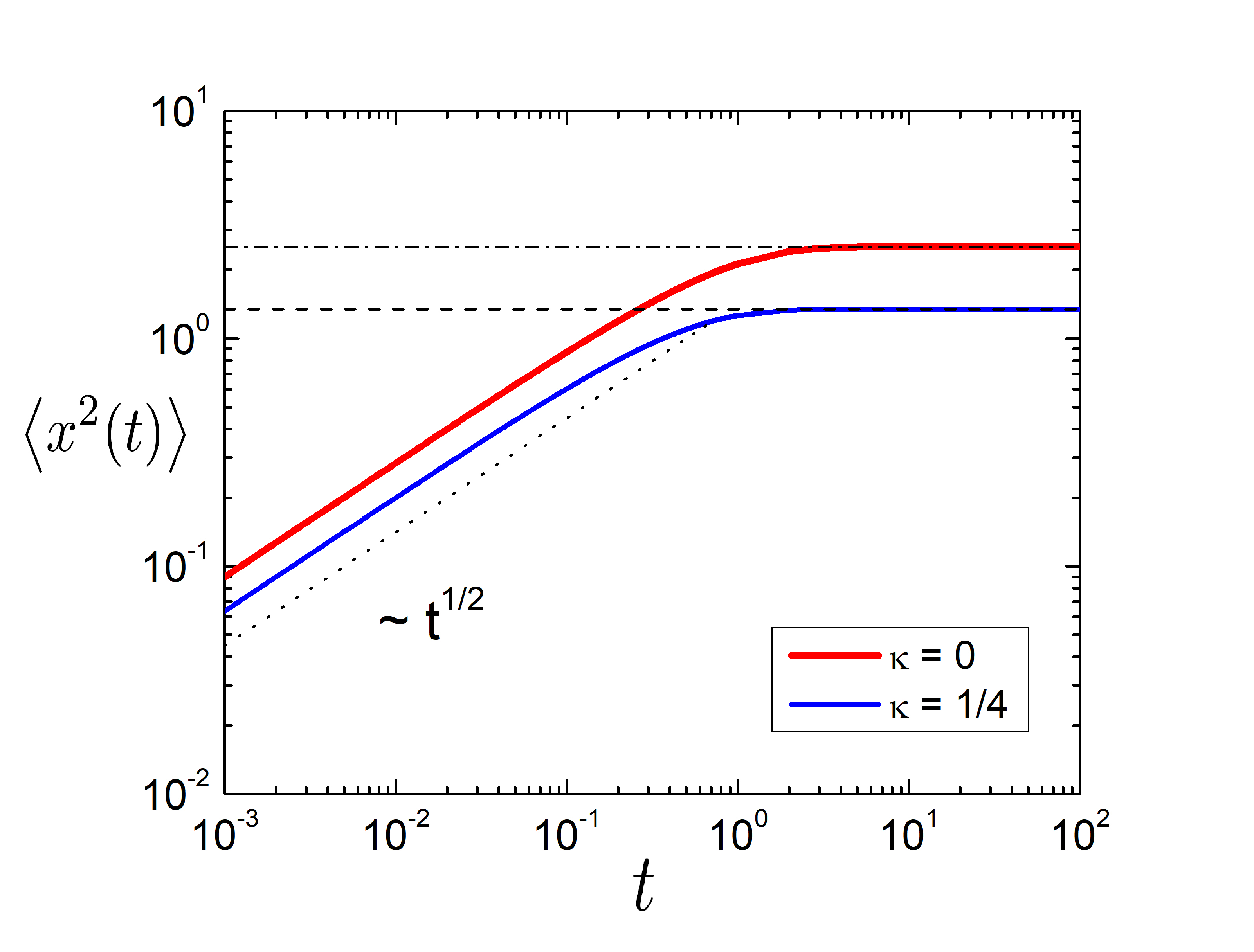}}\caption{The evolution of MSD with resetting for $r=1$, $\mu=\frac{4}{3}$, $\nu=4$, $\varkappa=\frac{1}{2}$, with and without confinement, $\kappa=\frac{1}{4}$ (Eq.~\ref{msdrescon}) and $\kappa=0$ (Eq.~\ref{msdresfree}), respectively. At short time it evolves according to $\sqrt{t}$ as in the free state. At $t\to \infty$ the MSD tends to the stationary values, given by Eqs. (\ref{resshort}) and (\ref{reslong}) for $\kappa=0$ (dashed dotted line) and $\kappa=\frac{1}{4}$ (dashed line), respectively. } \label{Gx2res}
\end{figure}

\subsubsection{Free MSD}

Let us examine the MSD without an external potential ($\kappa=0$).
Introducing Eq.~(\ref{x2k}) and Eq.~(\ref{surv}) into Eq.~(\ref{x2beg}), we obtain
\begin{eqnarray}\nonumber
\left\langle x^2(t)\right\rangle = K_{\alpha}e^{-rt}t^{\alpha}+
K_{\alpha}\int_0^t dt^{\prime} re^{-r(t-t^{\prime})}\left(t-t^{\prime}\right)^{\alpha}\,.\\\label{msdresfree}
\end{eqnarray} 
This integral can be calculated numerically, and the result is indicated by the red line in Fig.~\ref{Gx2res}. We now provide analytical estimations. Integration at long times $t\to\infty$ yields
\begin{eqnarray}\label{x21}
\left\langle x^2(t)\right\rangle=K_{\alpha}r^{-\alpha}\gamma\left(\alpha+1, rt\right)\,,
\end{eqnarray}
where $\gamma(a,z)$ is the lower incomplete Gamma function. The MSD rapidly reaches a steady state
\begin{eqnarray}
\left\langle x^2(t)\right\rangle_s = K_{\alpha}r^{-\alpha}\Gamma\left(\alpha+1\right)\,.
\end{eqnarray}
\blue{This is in agreement with the general fact that any process with a power-law time dependence of the MSD under Poissonian resetting reaches a steady state \cite{msdgen}}.
Introduction of Eqs.~(\ref{Ka}-\ref{anu}) into the last expression yields
\begin{eqnarray}\label{resshort}
\left\langle x^2(t)\right\rangle_s =\frac{\Gamma\left(\frac{3}{\mu}\right)}{\Gamma\left(\frac{1}{\mu}\right)}\frac{\mu^{\frac{2}{\mu}}\left(\nu\varkappa\right)^{\frac{2}{\nu}}}{r^{\frac{2}{\nu}}}\Gamma\left(\frac{2}{\nu}+1\right)\,.
\end{eqnarray}
This value of the stationary MSD is depicted by the dashed-dotted line in Fig.~\ref{Gx2res}.
The case $\mu=2$ corresponds to the resetting of the standard free Brownian motion, and the MSD attains the steady-state value
 \begin{eqnarray}
\left\langle x^2(t)\right\rangle_s =\frac{2\varkappa}{r}\,.
\end{eqnarray}

\begin{figure}\centerline{\includegraphics[width=0.5\textwidth]{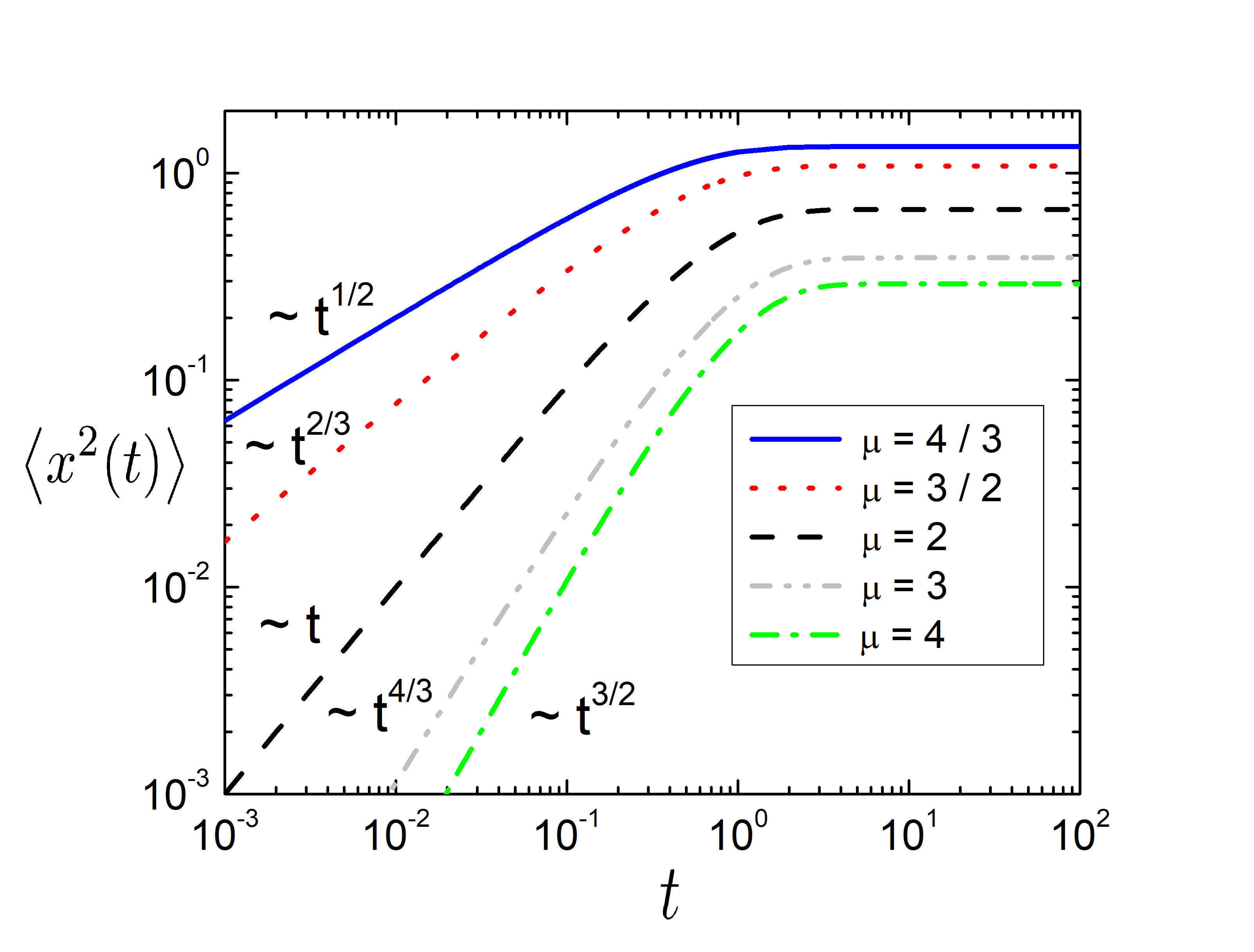}}\caption{The evolution of MSD with resetting for $r=1$, $\mu=\frac{4}{3},\, \frac{3}{2},\, 2,\, 3,\, 4$, $\varkappa=\frac{1}{2}$, $\kappa=\frac{1}{4}$ (Eq.~\ref{msdrescon}). The time-dependence at short times corresponds to free diffusion without resetting, given by Eqs.~(\ref{x2k}-\ref{anu}). The case $\mu=2$ corresponds to ordinary Brownian motion in confinement. } \label{Gx2resmu}
\end{figure}

\subsubsection{MSD in confinement}

Now we investigate the MSD in the presence of an external potential ($\kappa\ne 0$).
Introducing Eq.~(\ref{x2full}) into Eq.~(\ref{x2beg}) leads to
\begin{eqnarray}\nonumber
\left\langle x^2(t)\right\rangle=\frac{\Gamma\left(\frac{3}{\mu}\right)\mu^{\frac{2}{\mu}}\varkappa^{\frac{2}{\nu}}}{\Gamma\left(\frac{1}{\mu}\right)}\left(\frac{1-e^{-\nu\kappa t}}{\kappa}\right)^{\frac{2}{\nu}}e^{-rt}\\+r\frac{\Gamma\left(\frac{3}{\mu}\right)\mu^{\frac{2}{\mu}}\varkappa^{\frac{2}{\nu}}}{\Gamma\left(\frac{1}{\mu}\right)\kappa^{\frac{2}{\nu}}}\int_0^t \left(1-e^{-\nu\kappa \tau}\right)^{\frac{2}{\nu}}e^{-r\tau}d\tau\,.\label{msdrescon}
\end{eqnarray} 
Here $\tau=t-t^{\prime}$. The numerical integration of this expression is indicated by the solid blue line in Fig.~\ref{Gx2res}. The evolution of the MSD for different values of $\mu$ is shown in Fig.~\ref{Gx2resmu}. As $\mu$ increases, the slope at earlier times becomes more abrupt, and the stationary state is obtained earlier and approaches to zero. \red{The behavior of the MSD at short times $rt\ll 1$ is the same as that without resetting (Eqs.~\ref{x2k}-\ref{Ka}), because the number of resetting events at short times is negligible.} %$\mu>2$ corresponds to superdiffusion and $1<\mu<2$ corresponds to subdiffusion. The value $\mu=2$ corresponds to standard Brownian motion in confinement. In this case, at short times, the MSD exhibit a linear time-dependence.

Let us determine the stationary value of the MSD. Substituting $\xi=e^{-r\tau}$ into Eq.~(\ref{msdrescon}), and neglecting the first term yields
\begin{eqnarray}
\left\langle x^2(t)\right\rangle=\frac{\Gamma\left(\frac{3}{\mu}\right)\mu^{\frac{2}{\mu}}\varkappa^{\frac{2}{\nu}}}{\Gamma\left(\frac{1}{\mu}\right)\kappa^{\frac{2}{\nu}}}\int_{e^{-rt}}^1 d\xi\left(1-\xi^{\frac{\nu\kappa}{r}}\right)^{\frac{2}{\nu}}\,.
\end{eqnarray} 
In the limit $t\to\infty$ the lower integration limit tends to zero. Using
\begin{equation}
\int_0^1\left(1-x^b\right)^a dx=\frac{\Gamma(1+a)\Gamma\left(1+\frac{1}{b}\right)}{\Gamma\left(1+a+\frac{1}{b}\right)}
\end{equation}
we obtain 
\begin{equation}\label{reslong}
\left\langle x^2(t)\right\rangle_s=\frac{\Gamma\left(\frac{3}{\mu}\right)\mu^{\frac{2}{\mu}}\varkappa^{\frac{2}{\nu}}}{\Gamma\left(\frac{1}{\mu}\right)\kappa^{\frac{2}{\nu}}}\frac{\Gamma\left(1+\frac{2}{\nu}\right)\Gamma\left(1+\frac{r}{\nu\kappa}\right)}{\Gamma\left(1+\frac{2}{\nu}+\frac{r}{\nu\kappa}\right)}\,.
\end{equation}
This steady state MSD is depicted by the dashed line in Fig.~\ref{Gx2res}.
%The MSD in confinement under resetting can be expressed trough the MSD without confinement, Eq.~(\ref{resshort}), as
%\begin{equation}\left\langle X^2(t)\right\rangle=\left\langle X^2(t)\right\rangle_s\left(\frac{r}{\nu\kappa}\right)^{\frac{2}{\nu}}\frac{\Gamma\left(1+\frac{r}{\nu\kappa}\right)}{\Gamma\left(1+\frac{2}{\nu}+\frac{r}{\nu\kappa}\right)}\end{equation}
The particle in confinement is located closer to the origin than the free particle.

For $\mu=2$ the MSD, given by Eq.~(\ref{reslong}), becomes
\begin{equation}
\left\langle x^2(t)\right\rangle_s=\frac{\varkappa}{\kappa+\frac{r}{2}}\,.
\end{equation}
In this case, the presence of both the external potential and resetting affects the behavior of the MSD in a similar way: the MSD has the same functional dependence on the resetting rate and potential strength.

\begin{figure}\centerline{\includegraphics[width=0.5\textwidth]{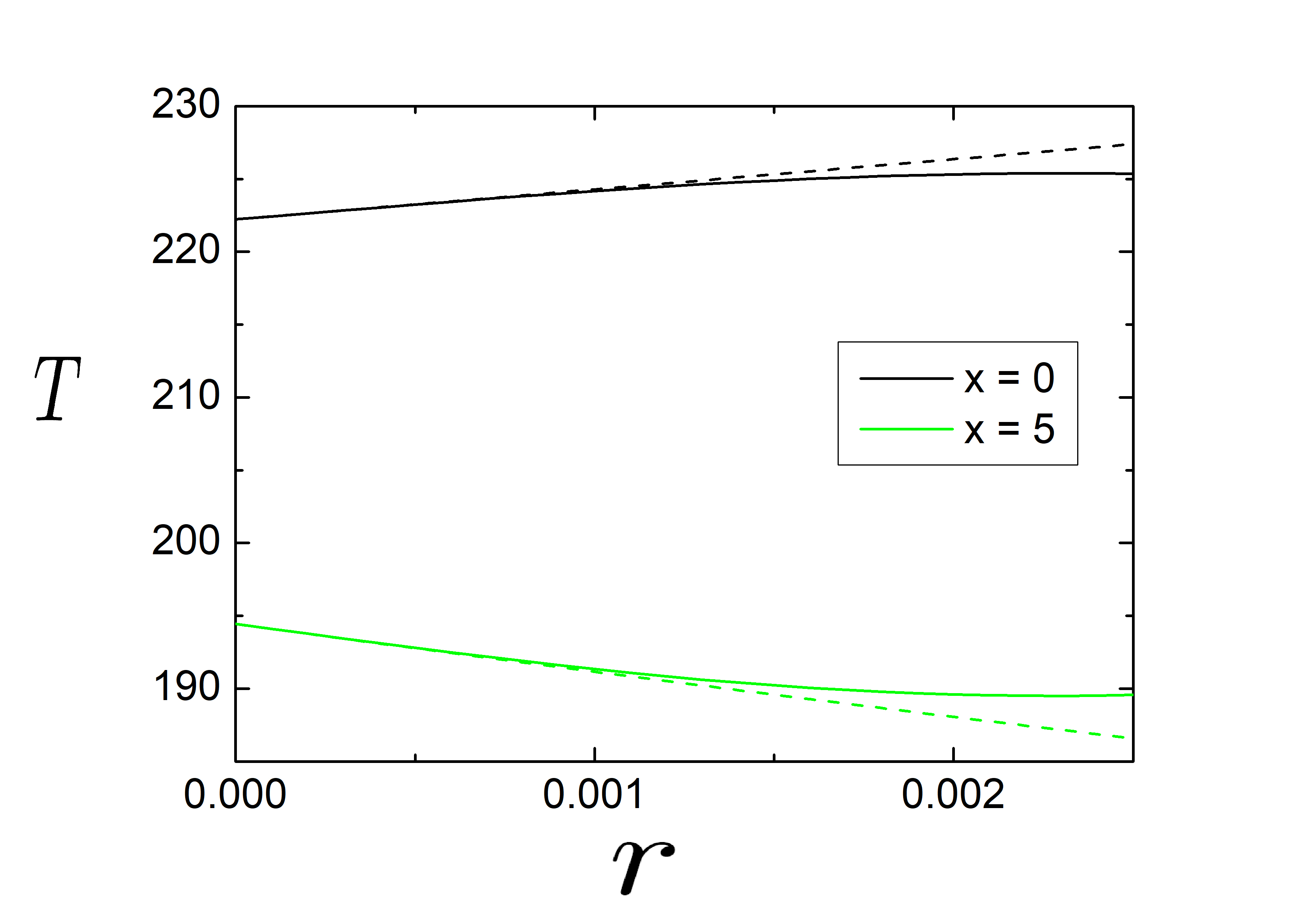}}\caption{\blue{Mean exit times $T$ for $b=10$, $x=0$ (black lines) and $x=5$ (green lines). In the first case, $T$ increases with increasing of $r$, which makes resetting unbeneficial to the search process. In the latter case, resetting favors the search process. $T$ decreases with increasing of $r$ for $r\ll 1$, then it eventually reaches its minimum at some optimal resetting rate $r^*$ and grows with increasing of $r$ for $r>r^*$. At $r\to\infty$, $T\to\infty$,  because in this case, the particle always remains in the vicinity of the resetting position and never reaches the boundaries. The solid lines correspond to Eq.~(\ref{Tfull}), the dashed lines to Eq.~(\ref{Tapprox})}. }\label{GTexit}
\end{figure}

\subsection{Mean first passage time}

Let us consider the survival probability $Q(t,y)$ in an interval  $[-b,b]$ 
with absorbing boundaries. 
If the particle reaches the end of the interval ($y=-b$ or $y=b$), it exits the diffusion process. Let the particle begin its diffusion at time $s=0$ 
from position $y$ and reach the boundary at time $t$. Then the survival probability can derived by performing the integration
\begin{equation}
Q(t,y)=\int^{b}_{-b}f(t,x|y)\,dx .
\end{equation}
 The classical backward Fokker-Planck equation for survival probability has the form \cite{PRLinitial, palprasad}
\begin{equation}
\frac{\partial Q(t,y)}{\partial t}=
\varkappa\frac{\partial^2Q(t,y)}{\partial y^2}-rQ(t,y)+rQ(t,x) .
\label{fmu32}
\end{equation}
with boundary  
$$
Q(t,-b)=Q(t,b)=0
$$ 
and initial 
$$
Q(0,y)=1,
$$ 
conditions.

%where the variable $x$ is now the initial position in backward equation.
For the generalized case (parameter values $n=2$, $\mu=3/2$, $\kappa=0$, $y\geq 0$), the following higher-order backward Fokker-Planck equation can be obtained:
%Let us write down the Fokker-Planck equation for PDF with resetting: 
\begin{eqnarray}\nonumber
\frac{\partial Q(t,y)}{\partial t}&=&\varkappa\left(-\frac{\partial}{\partial y}\frac{1}{2 y^2}-\frac{\partial^{2}}{\partial y^{2}}\frac{1}{2 y}+\frac{\partial^{3}}{\partial y^{3}}\right)Q(t,y)\\
&-&rQ(t,y)+rQ\left(t,x\right)\,.
\label{fmu32}
\end{eqnarray}
For simplicity, below we use $\varkappa=1$.
%Survival probability $Q(t,y)$ of a particle  
%\begin{equation}
%Q(x,t)=\int_{-b}^{b} f(\xi,t)d\xi
%\end{equation}
%is the probability that the particle has hit neither of the boundaries until time $T$ or survives within the interval $[-b,b]$ until time $t$.

The mean exit time for resetting of Brownian particles in the interval with absorbing boundaries can be expressed as the integral of survival probability:
\begin{equation}
T(y)=\int_0^{\infty} Q(t,y)\,dt.
\end{equation}
$T(y)$ for Brownian particles with resetting in the interval with absorbing boundaries was considered in \cite{palprasad}. 
 
%We assume that the particle returns to the position $-b<x<b$ during the resetting events. 

By integrating Eq.~(\ref{fmu32}), we obtain
\begin{equation}
\left(-\frac{\partial}{\partial y}\frac{1}{2 y^2}-\frac{\partial^{2}}{\partial y^{2}}\frac{1}{2y}+\frac{\partial^{3}}{\partial y^{3}}\right)T(y)=-1+rT(y)-rT(x)\,.
\label{Tmain}
\end{equation}
with absorbing boundary conditions 
\red{\begin{equation}T(-b)=T(b)=0.\end{equation}}
Without resetting ($r=0$), the solution of Eq.~(\ref{Tmain}) is \cite{Ser25}:
\begin{equation}\label{T0}
T_0(y)=\frac{2}{9}\left(b^3-|y|^3\right)\,.
\end{equation}
We now consider the initial position to be $y$ (which is a variable), while keeping the resetting position $x$ fixed. We solve the equations, and at the end set $y=x$. Because of the symmetry of the task, we provide a derivation for $y\ge 0$. 
The case $y<0$ can be considered analogously. 
We consider the case of a small resetting rate and derive $T(y)$ in terms of perturbation theory as a polynomial in the restart rate \cite{pallandau, palbook}:
\begin{equation}\label{Tper}
T(y)=T_0(y)+rT_1(y)+r^2T_2(y)+...
\end{equation}
Introducing Eq.~(\ref{Tper}) and Eq.~(\ref{T0}) into Eq.~(\ref{Tmain}) and collecting the terms proportional to $r$, we obtain
%\begin{eqnarray}\nonumber\left(\frac{\partial}{\partial x}\frac{1}{2 x^2}+\frac{\partial^{2}}{\partial x^{2}}\frac{1}{2 x}-\frac{\partial^{3}}{\partial x^{3}}\right)T_1(x)=T_0(x_0)-T_0(x)\\\label{T1}\end{eqnarray}
%Introducing Eq.~(\ref{T0}) into Eq.~\ref{T1} and performing the integration, we obtain
\begin{eqnarray}\label{eT1}
\frac{d^2T_1}{d^2y}-\frac{1}{2y}\frac{dT_1}{dy}=\frac{2}{9}b^3y-\frac{1}{18}y^4-T(x)y\,.
\end{eqnarray}
Solution of this differential equation is
\begin{eqnarray}\nonumber
T_1(y)=\frac{2}{9}T(x)\left(b^3-y^3\right)+\frac{1}{486}\left(b^3-y^3\right)\left(y^3-23b^3\right)\label{T1x}
\end{eqnarray}
%Introducing this expression into Eq.~(\ref{Tper}) and setting $T(x)=T(x_0)$, we obtain
%Now we introduce this expression into the expansion of $T(x)$ in the first approximation and set $T(x_0)=T(x)$, which means that the particle always starts its motion from the resetting point
%\begin{equation}\label{Tx0r}T(x_0)=\frac{2}{9}\left(b^3-x^3\right)+rT_1(x_0)\end{equation}
%Now we can express $T(x_0)$ from this equation and get
%\begin{equation}T(x_0)=\frac{\frac{2}{9}\left(b^3-x_0^3\right)+\frac{1}{486}\left(x_0^3-23b^3\right)\left(b^3-x_0^3\right)r}{1-\frac{2}{9}\left(b^3-x_0^3\right)r}\end{equation}
%\begin{equation}T(x_0)=\frac{2}{9}\left(b^3-x_0^3\right)\times\frac{1+r\left(x_0^3-23b^3\right)/108}{1-2r\left(b^3-x_0^3\right)/9}\end{equation}
%\begin{equation}T(x_0)=T_0(x_0)\frac{1+r\left(x_0^3-23b^3\right)/108}{1-2r\left(b^3-x_0^3\right)/9}\end{equation}
To derive $T_2(y)$ we insert Eq.~(\ref{Tper}) into Eq.~(\ref{Tmain}) and collect terms proportional to $r^2$:
\begin{eqnarray}\label{eT2}
\frac{d^2T_2}{dy^2}-\frac{1}{2y}\frac{dT_2}{dy}=\int T_1(y)dy\,.
\end{eqnarray}
Performing the integration and solving the differential equation, one can get
%\begin{eqnarray}\nonumberT_2(x)&=&T(x_0)\left(\frac{4}{81} b^3 x^3-\frac{2}{9}x^3-\frac{1}{486} x^6\right)+\\&+&\frac{1}{2187}b^3 x^6-\frac{23}{2187}b^6 x^3-\frac{1}{229635}x^9\label{T2x}\end{eqnarray}
\begin{eqnarray}\nonumber
&&T_2(y)=\frac{2}{9}T(x)\left(\frac{2}{9} b^3 y^3-\frac{1}{108} y^6\right)\\
&&+\frac{1}{2187}b^3 y^6-\frac{23}{2187}b^6 y^3-\frac{1}{229635}y^9+c_2\,.\label{T2x}
\end{eqnarray}
The constant $c_2$ can be obtained by considering the boundary condition $T(b)=0$. Introducing Eq.~(\ref{T0}), Eq.~(\ref{T1x}) and Eq.~(\ref{T2x}) into Eq.~(\ref{Tper}) and formally setting $T(y)=T(x)$ and $y=x$, we obtain the mean exit time, when the particles initially start their motion from the resetting point: %The final expression has the following form
%\begin{widetext}\begin{eqnarray}T(y)=\frac{\frac{2}{9}\left(b^3-|y|^3\right)+r\frac{1}{486}\left(b^3-|y|^3\right)\left(|y|^3-23b^3\right)+\frac{1}{229635}r^2\left(2311 b^9 - 2415 b^6 |y|^3 + 105 b^3 y^6 - |y|^9\right)}{1-\frac{2}{9}r\left(b^3-|y|^3\right)+\frac{1}{486}r^2\left(23 b^6 - 24 b^3 |y|^3 + y^6\right)}\label{Tfull}\end{eqnarray}\end{widetext}
\begin{widetext}\begin{eqnarray}T(x)=\frac{\frac{2}{9}\left(b^3-|x|^3\right)+r\frac{1}{486}\left(b^3-|x|^3\right)\left(|x|^3-23b^3\right)+\frac{1}{229635}r^2\left(2311 b^9 - 2415 b^6 |x|^3 + 105 b^3 x^6 - |x|^9\right)}{1-\frac{2}{9}r\left(b^3-|x|^3\right)+\frac{1}{486}r^2\left(23 b^6 - 24 b^3 |x|^3 + x^6\right)}\label{Tfull}\end{eqnarray}\end{widetext}
By expanding this expression in series, we obtain the mean exit time in the second approximation with respect to the resetting rate $r$
%\begin{eqnarray}\nonumber&&T(y)=\frac{2}{9}b^3\left(1-|u|^3\right)+\frac{1}{486}b^6\left(1-24|u|^3+23 u^6\right)r+\\\nonumber&&\frac{1}{229635}b^9\left(1 - 105 |u|^3 + 2415 u^6 - 2311 |u|^9\right)r^2+O(r^3).\\\label{Tapprox}\end{eqnarray}
\begin{eqnarray}\nonumber&&T(u)=\frac{2}{9}b^3\left(1-|u|^3\right)+\frac{1}{486}b^6\left(1-24|u|^3+23 u^6\right)r+\\\nonumber&&\frac{1}{229635}b^9\left(1 - 105 |u|^3 + 2415 u^6 - 2311 |u|^9\right)r^2+O(r^3).\\\label{Tapprox}\end{eqnarray}
\red{Here, we introduced the dimensionless parameter $u=x/b$, where $|u|<1$.}
The dependence of the mean exit time on the resetting rates is shown in Fig.~\ref{GTexit}. If the initial position is located near the center, the mean exit time increases with the resetting rate. In this case, resetting is not beneficial to the search process. If it is located close to the boundary, the mean exit time decreases with the resetting rate and eventually reaches a minimum value. In this case, resetting favors the search. At large resetting rates, $T$ tends to infinity because, in this case the particle always remains in the vicinity of the resetting position and never reaches the boundaries. % In this case, the particle does not have time to move and is always located in the vicinity of the resetting point $x_0$.  

\red{It has been shown that the sufficient condition for resetting to favor the search process is $CV>1$, where $CV$ is the coefficient of variation \cite{shlomimfpt}. $CV$ is equal to the ratio between the standard deviation and the mean first passage time of the underlying (without restart) first-passage time process. Similar to the case of a Brownian particle in a one-dimensional box \cite{palprasad}, one can show that $CV>1$ can be obtained if $T_1(y)<0$ or }
%\red{\begin{eqnarray}b^6-24 b^3 |x_0|^3+23 x_0^6<0\end{eqnarray}}
\red{\begin{eqnarray}1-24 |u|^3+23 u^6<0\end{eqnarray}}
\red{Solving this inequality, we can obtain a sufficient criterion for restart to be beneficial:}
\red{\begin{eqnarray}|u|>u^*=23^{-1/3}\approx 0.35\end{eqnarray}}

%The coefficient $T_2(x_0)<0$ if $0.241<u<0.314$.

\section{Conclusion}

We investigated the resetting of particles performing the motion, which can be described in terms of the generalized Ornstein-Uhlenbeck distribution. The PDF and MSD of both free diffusive motion and motion in confinement were considered. It was shown that the resulting PDF and MSD become stationary in both cases at long times. First, the NESS was established in the core region of the PDF. The MSD increased with an increase in the diffusion coefficient and decreased with the potential strength and resetting rate. The PDF exhibited an exponential dependence on the coordinates and became narrower in the presence of confinement. We also derived the mean exit time for a particle confined in a one-dimensional box with absorbing boundaries. Depending on the position of the resetting point $x$ with respect to boundary $b$, resetting may either favor or prohibit the search process. \red{The restart is beneficial if the particle starts closer to either of the boundaries, $|u|=|x|/b>u^*$. If the particle starts in the region, centered around the origin, $|u|<u^*$, restart only prolongs completion.}

\section*{Acknowledgments} 

We thank the anonymous reviewer for their insightful
remarks.

\end{document}